# Shell model study of Neutron rich Oxygen Isotopes

**P.C.Srivastava and I.Mehrotra**
Nuclear Physics Group, Deptt. Of Physics,
University of Allahabad, India.

**Abstract:** Shell model calculations for low lying energy states of neutron rich oxygen isotopes $^{19}$O, $^{20}$O, $^{21}$O, $^{22}$O, $^{23}$O, $^{24}$O, $^{25}$O, $^{26}$O have been performed using OXBASH code. The configuration space consists of $0d_{5/2}$, $1s_{1/2}$, $0d_{3/2}$ orbital for neutrons outside the $^{16}$O core. Different interactions namely, Wildenthal, Preedom-Wildenthal, Wildenthal-Mcgrory and renormalized Kuo and Brown, which are either empirical or realistic in nature have been used in the calculation, The calculated energy spectrum are in good agreement with the experimental data wherever available for the empirical interactions and the correct ordering of levels is reproduced. The levels obtained from realistic interactions, though have a small rms deviation, do not reproduce the correct ordering in some of the cases. In the case of $^{21}$O, realistic interactions predict a much too compressed $1/2^+$ state at energy 0.157 MeV compared to the experimental value of 1.218 MeV. For even isotopes the variation of the energy of the first $2^+$ exited state has been studied as a function of neutron number N. A sharp rise in the value is observed at N=16 for both empirical and realistic interactions and only at N=14 for empirical interactions. Significantly higher energy of first $2^+$ exited state compared to the value in the neighboring even-even nuclei is considered as a signature for magic nuclei at N=14 and N=16.

# Introduction:

The oxygen isotopes from A=16 to 28, are the heaviest nuclei for which neutron dripline has been experimentally established. It has been reported that nuclei up to $^{24}$O are inside and that of $^{25}$O, $^{26}$O, $^{27}$O and $^{28}$O are outside the dripline. Experimental data on the energy level of O isotopes up to excitation energy of the order 10MeV has been reported in the literature recently.

Existing theoretical studies in the literature are based on two approaches. In the first approach a renormalized N-N interaction is used in a limited configuration space. Following this line of approach Brown et al. has calculated energy levels of $^{22}$O and $^{24}$O with normalized Bonn-A interaction and phnomenogical one-boson exchange interaction. Experimental have studied $^{21}$O, by multi-nucleon transfer reactions are in good agreement with experimental values. Experiment has proposed shell closer for $^{22}$O for N=14, and 0p-0h configuration for $^{22}$O. These authors has assigned the level scheme of $^{20}$O and $^{21}$O to 0p-2h configurations and 0p-1h configurations respectively. Similarly the level scheme of $^{23}$O and $^{24}$O are assigned to 1p-0h and 2p-0h respectively.

In the second line of approach the two body matrix elements are treated as parameters, and their values are obtained from best fit to experimental data. Brown and coworkers have carried extensive studies of energy level and spectroscopic properties of sd-shell nuclei in terms of a unified Hamiltonian applied in full sd-shell model space. The universal Hamiltonian was obtained from a least square fit of 380 energy data with experimental errors of 0.2MeV or less from 66 nuclei. The USD Hamiltonian is defined by 63 sd-shell two body matrix element (TBME) and their single particle energies. In more recent work Brown and co-workers have modified USD type Hamiltonian to USDA and USDB based on updated set of binding energy and energy levels of O, F, Ne, Na, Mg and Si isotopes.

In the present work we have calculated the energy levels of $^{20}$O, $^{21}$O, $^{22}$O, $^{23}$O, $^{24}$O, $^{25}$O and $^{26}$O isotopes up to 7MeV using modified surface delta interaction in the model space of $0d_{5/2}$, $1s_{1/2}$, $0d_{5/2}$. We use the following standard definitions for the rms deviation.

We use the following standard definitions for the rms deviation,

$$\sigma = \{1/N \sum_i [E_{exp}(i) - E_{cal}(i)]^2\}^{1/2}$$

# Results and discussion:

The energy spectra of oxygen isotopes are as,

(a) $^{19}$O: In case of $^{19}$O, ordering of level is correctly reproduced by w, pw and sdm interaction while in kuosd interaction in place of 5/2+ state we observe 3/2+ state. The mean deviation is lowest for sdm interaction and is highest for kuosd interaction.

(b) $^{20}$O: In case of $^{20}$O, ordering of level is correctly reproduced by w, pw, and sdm interaction while in kuosd interaction in place of first 4$^+$ state we observe 2+ state. The mean deviation is lowest for sdm interaction and is highest for kuosd interaction. In case of kuosd interaction we also find bunching of state above 2 MeV. Preedom-wildenthal interaction gives more exact result in comparison to other interactions. The realistic G-matrix does not give the correct ordering but the excitation energy of first 2+ states is nearest to the experimental data.

(c) $^{21}$O: In case of $^{21}$O, ordering of energy level is correctly reproduced by w, pw, and sdm interaction, while in kuosd interaction ground state is 1/2 $^+$ instead of 5/2$^+$. The mean deviation is lowest for sdm interaction and it is highest for pw interaction. In case of pw interaction we find bunching of state above 4 MeV while bunching of state is found above 3MeV in case of kuosd interaction. The G-matrix interaction does not reproduce the correct ground state.

(d)    $^{22}$O: In all the cases the second 2+ state is obtained lower than 3+ state as against the experimental observation in which 3+ state lies lower than second 0+ state .The spectrum obtained with sdm and G-matrix are compressed compare to the experimental data in all the cases.

(e)    $^{23}$O: Its one neutron separation energy is 2.7MeV. No gamma rays expected in this experiment. So exited state in $^{23}$O exist it is likely to be 5/2+,which should be decay in ½+ ground state by an E2 transition, relative to the 0p-0h model for $^{22}$O,the $^{23}$O levels are of 1p-0h and 2p-ah character. Of these only the lowest1p-0h S1/2(1/2+) is predicted to be bound and this agrees with present experiment. The first excited 5/2+ state at 2.72 MeV is the lowest state and is dominated by 2p-1h configurations.

(f)    $^{24}$O: In $^{24}$O the neutron separation energy is 3.7MeV,the first exited state is predicted at 4.18MeV, This explain non transition of gamma in $^{24}$O. In case of sdm interaction the exited states become more closer between 3-5MeV. In case of pw interaction first excited state become more higher.

(g) $^{25}$O: In case of sdm interaction energy state become more closer while in case of pw interaction energy level become more spread.

(h) $^{26}$O: In case of sdm interaction energy levels become more compressed, while in pw interaction the levels become more spread.

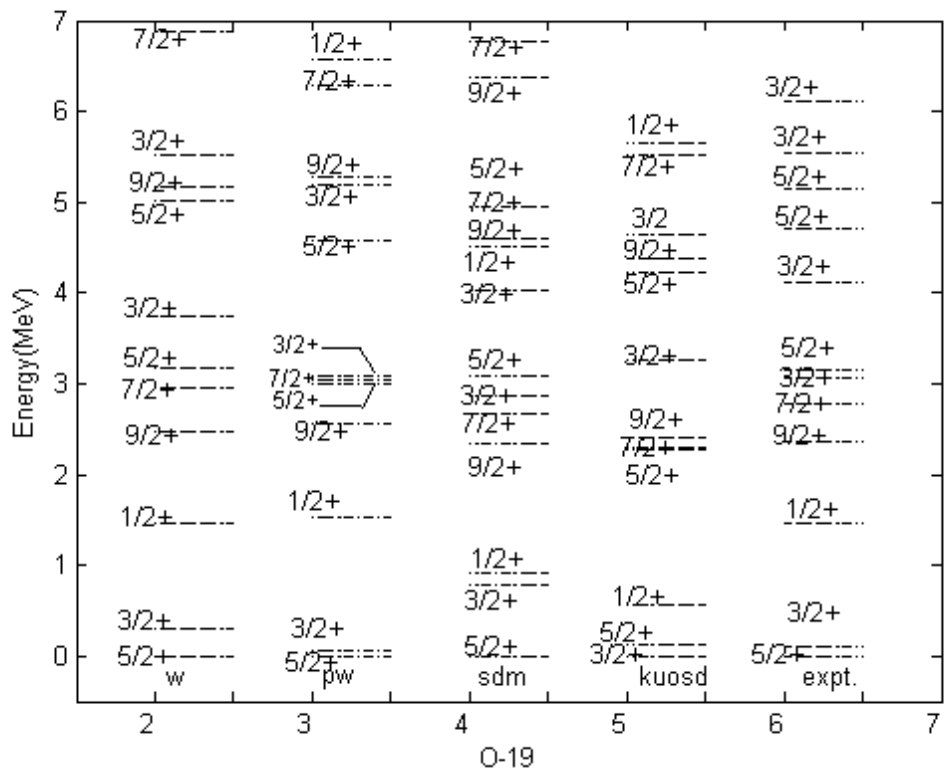

Fig. (1). Comparison of experimental and theoretical levels for $^{19}$O.

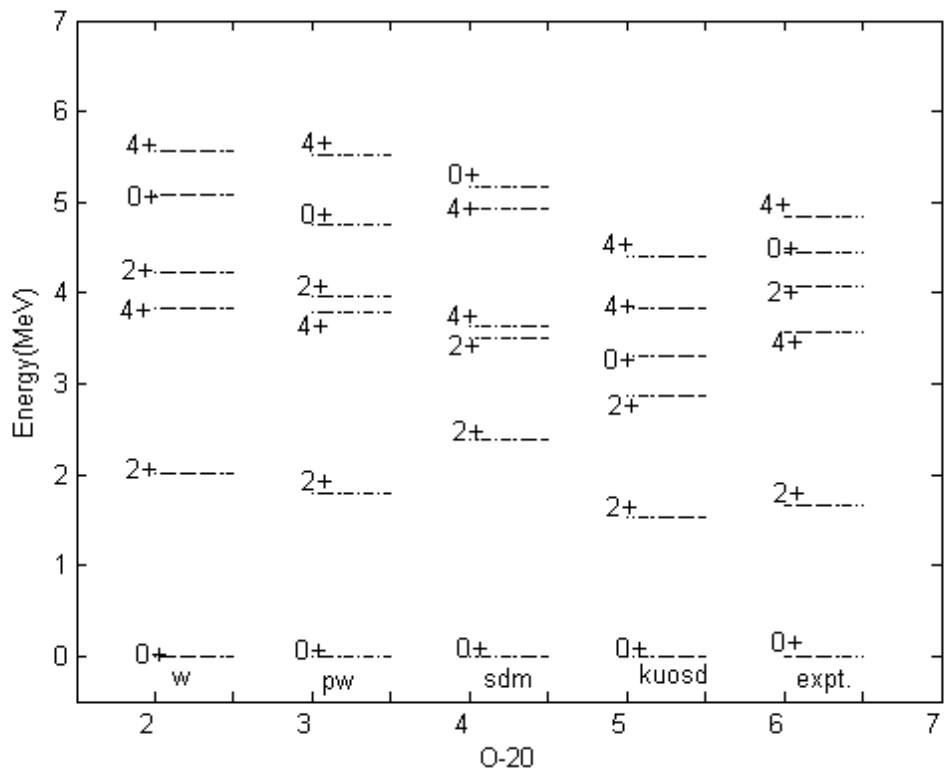

*Fig. (2). Comparison of experimental and theoretical levels for $^{20}O$.*

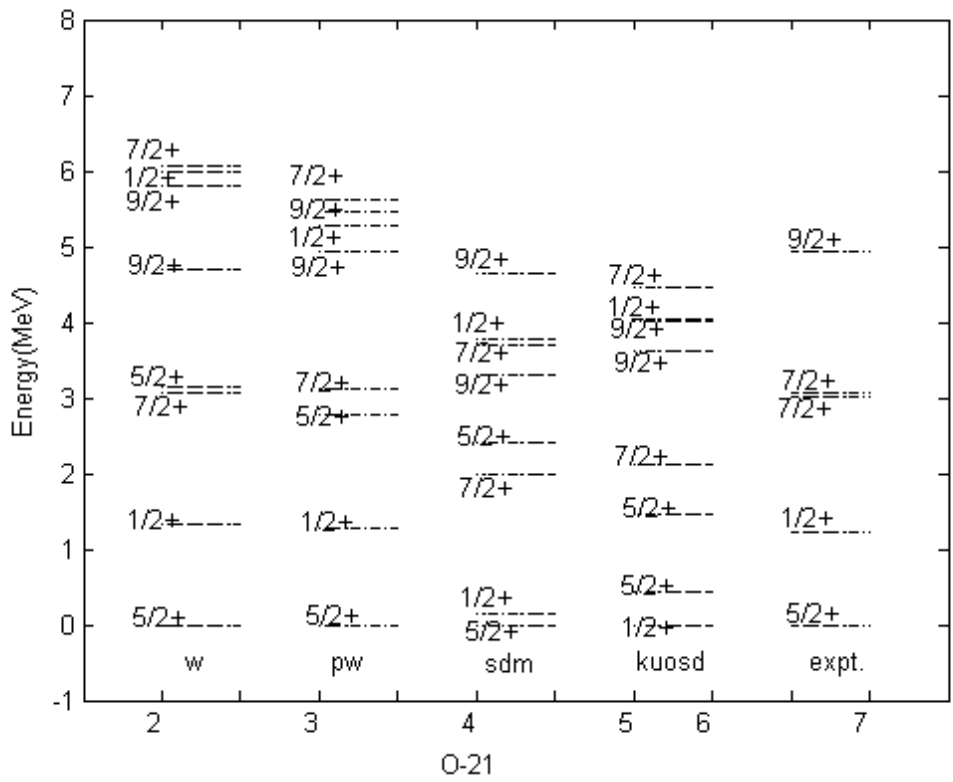

Fig. (3). Comparison of experimental and theoretical levels for $^{21}$O.

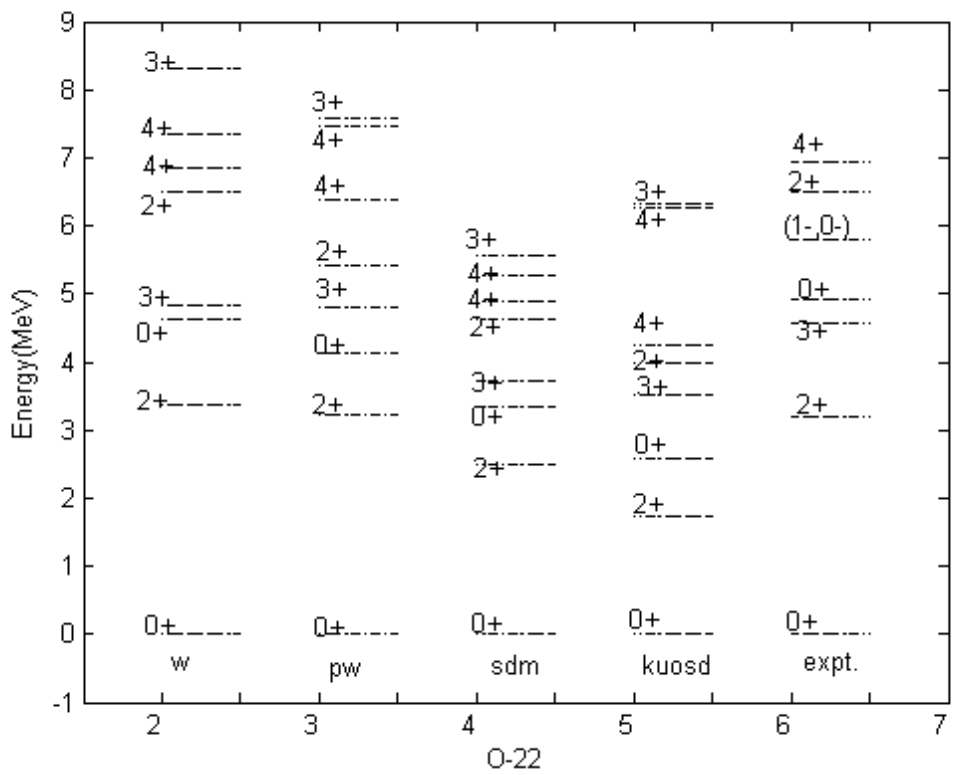

Fig. (4). Comparison of experimental and theoretical levels for $^{22}O$.

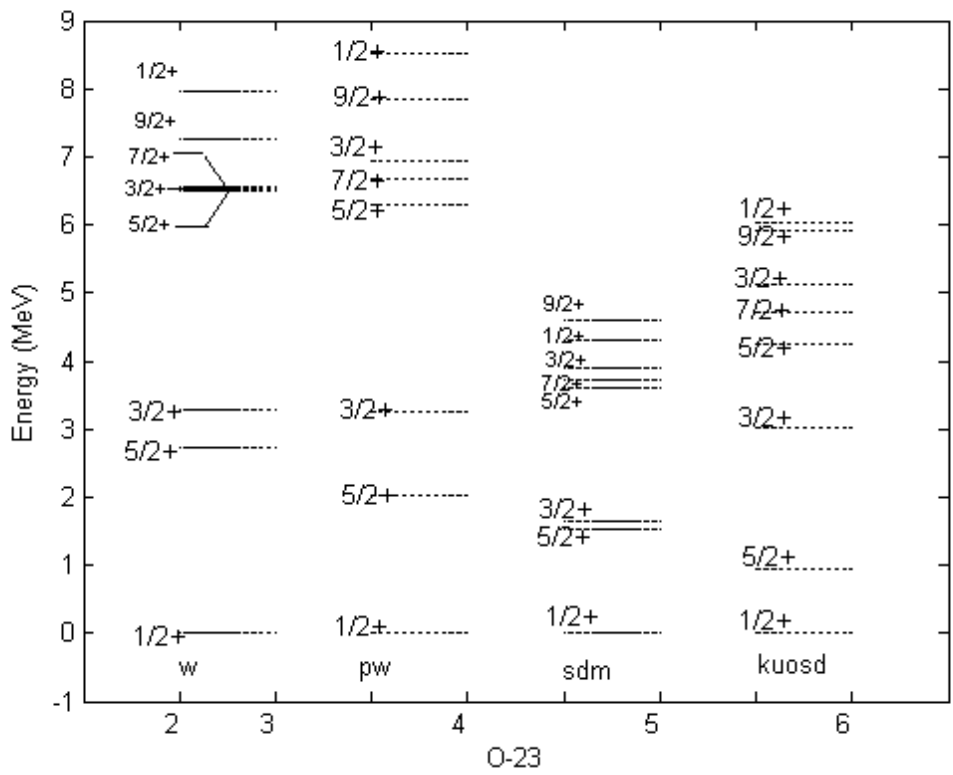

*Fig. (5). Comparison of experimental and theoretical levels for $^{23}$O.*

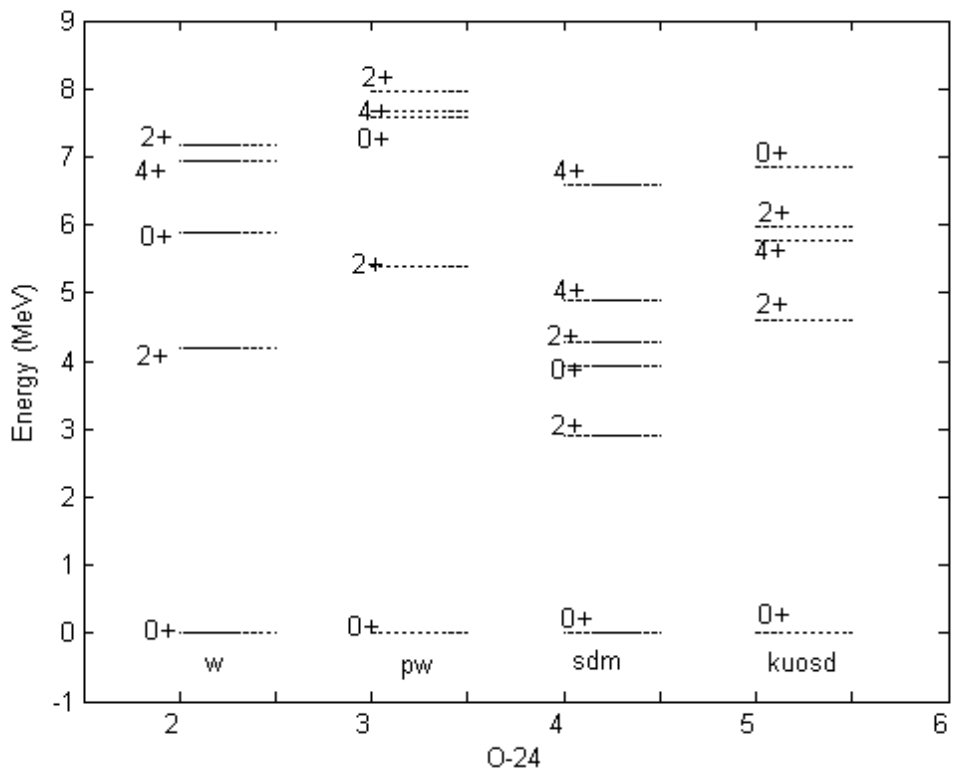

Fig. (6). Comparison of experimental and theoretical levels for $^{24}$O.

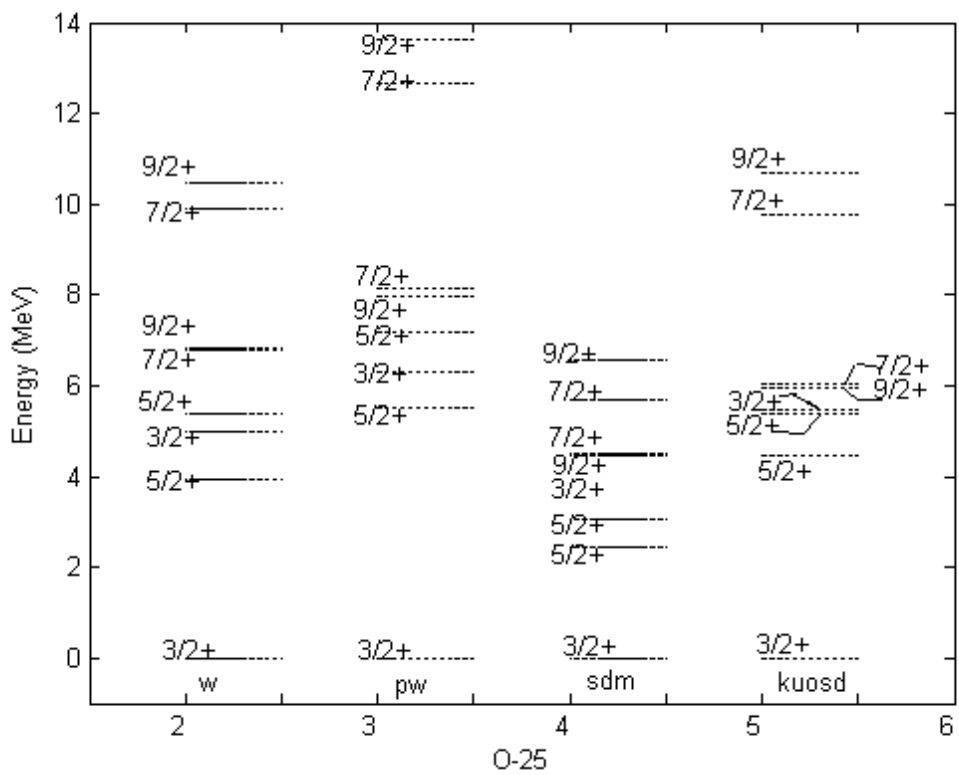

**Fig. (7).** Comparison of experimental and theoretical levels for $^{25}$O.

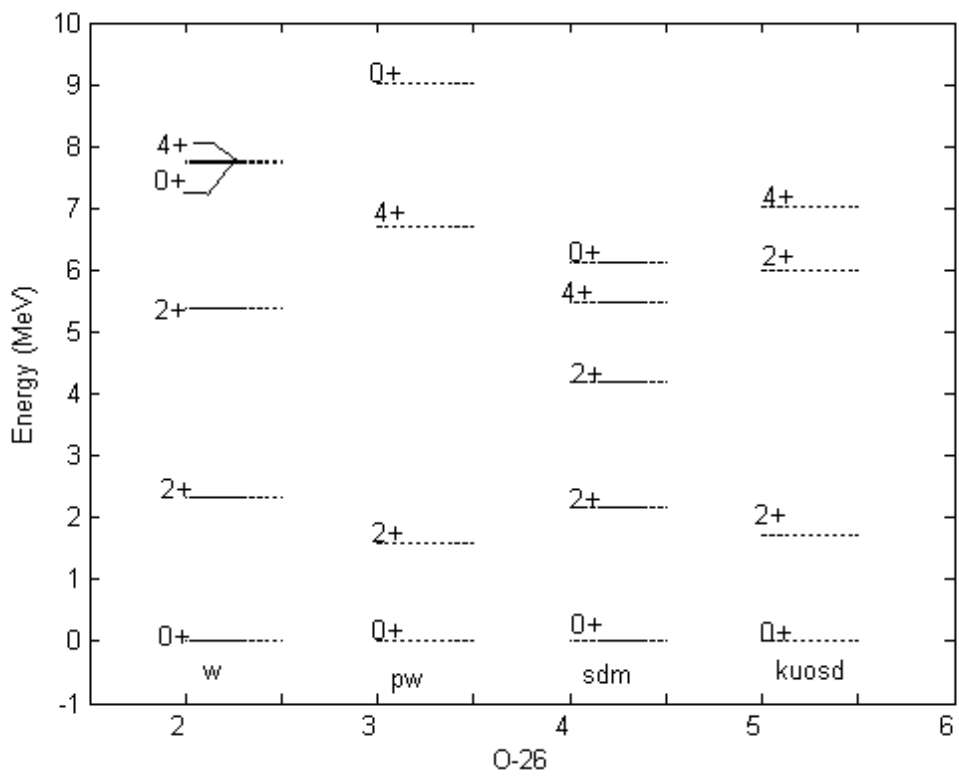

Fig. (8). Comparison of experimental and theoretical levels for $^{26}$O.

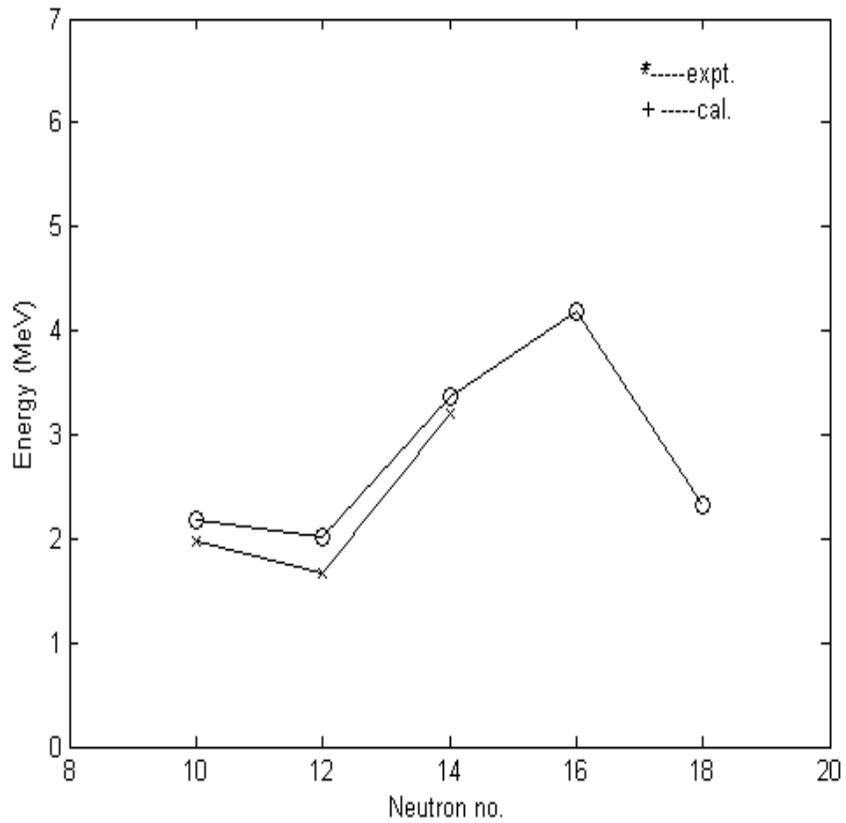

**Fig.(9). Excitation energy 2+ states in even-even nuclei, the experimental data (Astricks) and calculated data(open circles ) for w-interaction**

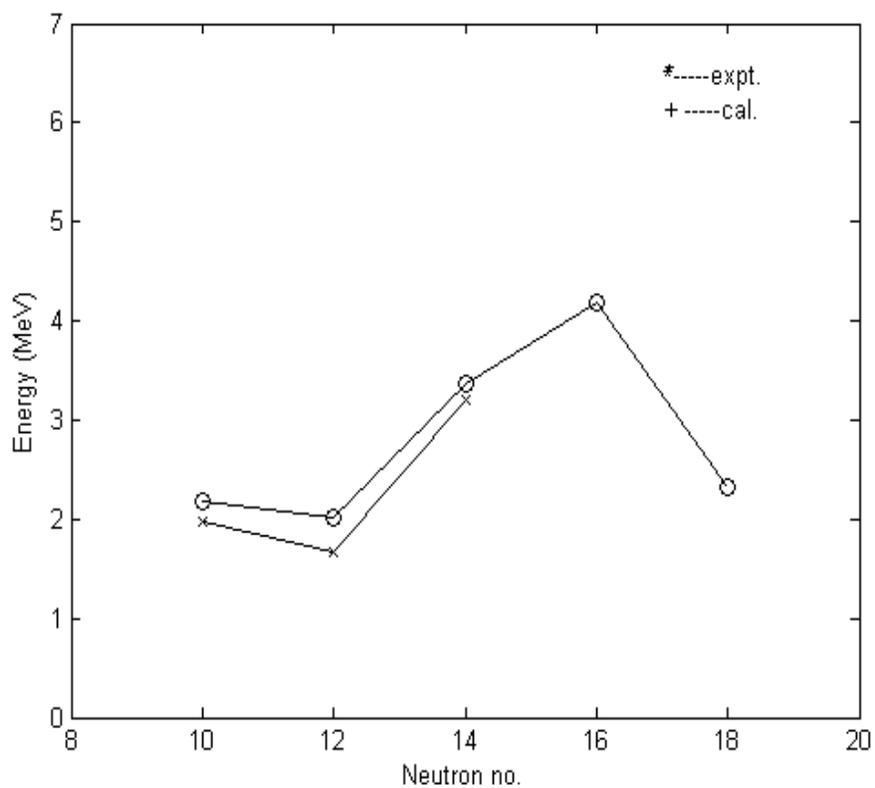

*Fig(10):* **Excitation energy 2+ states in even-even nuclei, the Experimental data (Astricks) and calculated data(open circles) for pw-interaction**

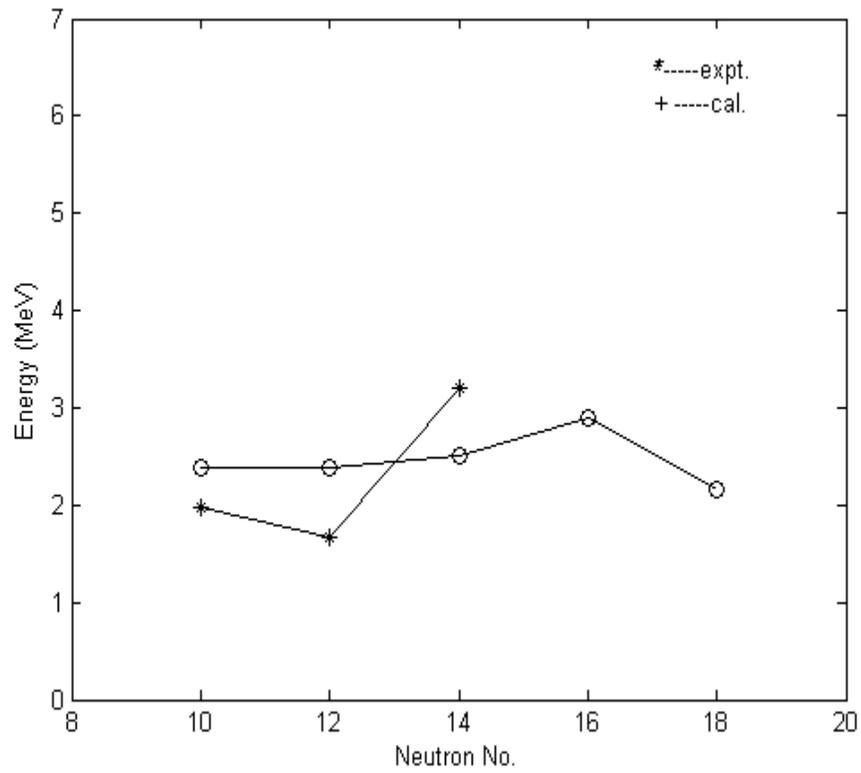

*Fig(11):* **Excitation energy 2+ states in even-even nuclei, the experimental**
  *data (Astricks) and calculated data(open circles) for* **sdm-interaction**

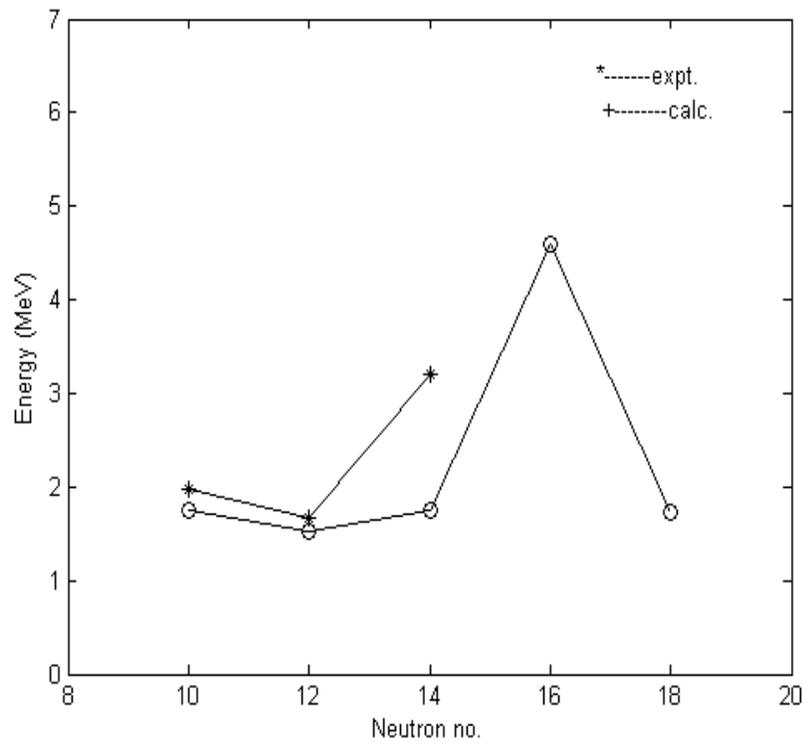

*Fig(12):* **Excitation energy 2+ states in even-even nuclei, the experimental** *data (Astricks) and calculated data(open circles) for* **kuoasd-interaction**

# REFERENCES:

[1] W.N. Catford et al., Nuclear Phys. A **503**, 63(1989).

[2] B.A. Brown et al., Phys. Rev. C **74**, 034315 (2006).

[3] B.A. Brown et al., Phys. Rev. C **72**, 057301 (2005).

[4] M. Stanoiu et al., Phys. Rev. C **69**, 034312 (2004).

[5] M.Thoennessen et al., Rep. Prog. Phys. **67**, 1187 (2004).

[6] Zhongzhou Ren et al., Phys.Rev.C **R20** (1995).

[7] D.Cortina-Gil et al., J. Phys. G: Nucl. Part.Phys.**31**, S1629 (2005).

[8] B.H.Wildenthal et al., Phys. Rev. C **4**, 1708 (1971).

[9] C.S. Sumithrarachchi et al., Phys. Rev. C **74**, 024322 (2004).

[10] E.K.Warburton et al., Phys. Rev. C **41**, 1147(1990).

[11] B.A.Brown et al., Ann. Phys. **182**,191(1988).

[12] M.Hjorth-Jensen et al., Phys. Rep. **261**,125(1995).

[13] B.H.Wildenthal, Prog. Part. Nucl. Phys.**11**, 5 (1984).

[14] B.A.Brown et al., Annu. Rev. Nucl. Part. Sci. **38**, 29 (1988).

[15] B.A.Brown's home page: http: // www.nscl.msu.edu /~brown/decay/

[16] E.Sauvan et al., Phys. Lett. B. **491** 1(2000).

[17] Experimental energy levels can be found from: www.nndc.bnl.gov/